\documentclass[prl,reprint, twocolumn,showpacs,preprintnumbers,amsmath,amssymb,nofootinbib,floatfix]{revtex4-1} 
\usepackage{graphicx}  
\usepackage{dcolumn}   
\usepackage{bm}        
\usepackage{amssymb}   
\usepackage{slashed}
\usepackage{color}
\usepackage{lipsum}
\usepackage{amsmath,amsthm,amsfonts,amssymb,amscd}
\usepackage{physics}

\begin{document}

\widetext

\title{CP violation induced by neutral meson mixing interference}

\author{Yin-Fa~Shen}\email{syf70280@hust.edu.cn}
\affiliation{School of physics, Huazhong University of Science and Technology, Wuhan 430074, China}
\author{Wen-Jie Song}\email{jaysong@hust.edu.cn}
\affiliation{School of physics, Huazhong University of Science and Technology, Wuhan 430074, China}
\author{Qin~Qin}\email{corresponding author: qqin@hust.edu.cn}
\affiliation{School of physics, Huazhong University of Science and Technology, Wuhan 430074, China}

\begin{abstract}
We propose a new kind of CP violation effect --- the double-mixing CP asymmetry --- in a type of cascade decays that involves at least two mixing neutral mesons in the decay chain. It is induced by the interference between different oscillation paths of the neutral mesons in the decay process. The double-mixing CP asymmetry is of critical importance for phenomenology, providing opportunities for clean determination of CKM phase angles free of uncertainties induced by the strong dynamics. To illustrate this point, we perform a phenomenological analysis on two examples: $B^0_s \to \rho^0 K \to \rho^0 (\pi^-\ell^+{\nu}_\ell)$ and $B^0 \to D^0 {K} \to D^0 (\pi^+\ell^-\bar{\nu}_\ell)$. Our results demonstrate that the double-mixing CP asymmetry can be numerically significant in the absence of strong phases, as shown by the former example. Additionally, the latter example showcases the direct extraction of weak and strong phases from data, without the need for theoretical inputs.
\end{abstract}


\maketitle


\textit{Introduction.}---The CP violation has always been playing a key role in particle physics. Measurements of CP violation effects in flavor processes are crucial to determine the Cabibbo-Kobayashi-Maskawa (CKM)~\cite{Cabibbo:1963yz,Kobayashi:1973fv} matrix, whose unitarity is a critical test of the standard model (SM). Moreover, the observed matter-antimatter asymmetry in the universe requires the CP violation as one of the criteria~\cite{Sakharov:1967dj}, but the CP violation in the SM is too small to be the only source~\cite{Bernreuther:2002uj,Canetti:2012zc}. Therefore, precision tests of CP asymmetries and searches for their non-standard source may open a window to physics beyond the SM.

In hadron decays, although the CP violation is induced by the weak interaction, its visualization typically requires interplay between weak and strong interactions and thus receives pollution from strong dynamics. Therefore, exploring CP asymmetries in diverse physical observables with diverse dependence on strong dynamics would be beneficial for unraveling the mystery of the CP violation~\cite{Yu:2017oky}. Particularly, observables free of strong phases provide a clean environment~\cite{BESIII:2021ypr,Wang:2022fih}. To this end, we propose a novel CP violation observable --- double-mixing CP asymmetry, which is induced by interferences between different mixing paths in a single cascade decay. It allows for the determination of weak phases avoiding strong-dynamics pollution in appropriate channels.

\begin{figure}[htbp]
\centering
\vspace{-0.2cm}
\includegraphics[width=1.0\linewidth]{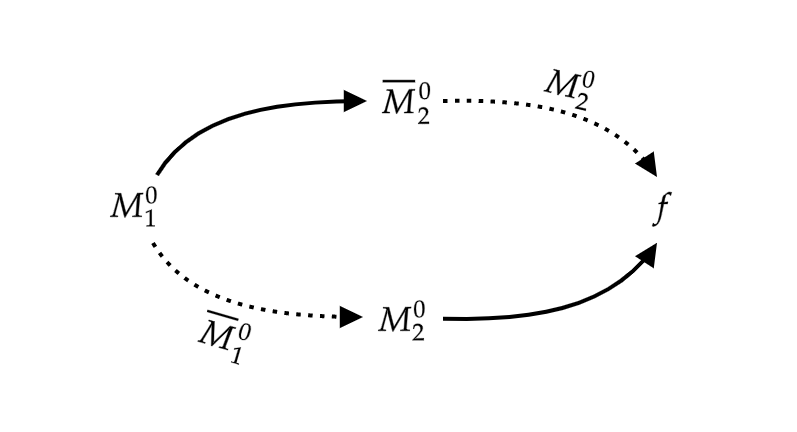}
\vspace{-1cm}
\caption{Interference between two oscillation paths in the cascade decay $M_1\to M_2\to f$. The decay products associated with $M_2$ are not displayed.}
\label{fig:2chain}
\end{figure}

The double-mixing CP asymmetry exists in cascade decay chains, in which at least two mixing neutral mesons are involved. Such a process typically starts from a primary neutral meson $M_1^0$, which decays, before or after oscillating to its antiparticle $\bar{M}_1^0$, into a secondary neutral meson $\bar{M}_2^0(M_2^0)$ and other particles. The secondary neutral meson $\bar{M}_2^0(M_2^0)$ also further decays, before or after oscillating, into directly detectable particles by detectors. The process $M_1\to M_2\to f$ (the particles produced associated with $M_2$ in the decay are omitted for simplicity) happens via multiple quantum paths, which interfere with each other. One example is shown in FIG.~\ref{fig:2chain}, which allows two oscillation paths $M_1^0\to \bar{M}_2^0 \to M_2^0$ and $M_1^0\to \bar{M}_1^0 \to M_2^0$. The interference between the two paths induces the double-mixing CP asymmetry, which is our focus in this study. There also exist other interfering paths such as $M_1^0\to \bar{M}_1^0 \to M_2^0\to \bar{M}_2^0$ and $M_1^0\to \bar{M}_2^0$. In cases where $M_2^0$ and $\bar{M}_2^0$ decay into the same final state, usually a CP eigenstate, there are interferences between the four paths, and induce more fruitful CP violation observables. 

The double-mixing CP asymmetry possesses a distinctive phenomenological significance. Its dependence on two time variables, the oscillation time $t_1$ of $M_1^0$ and $t_2$ of ${M}_2^0$, allows for a two-dimensional time-dependent analysis on $M_1^0(t_1)\to \bar{M}_2^0(t_2)\to f$, making it a new measurement tool. Practically, the double-mixing CP asymmetry can be numerically very significant in certain decay channels, to be discovered and measured at flavor experiments such as BESIII~\cite{BESIII:2020nme}, Belle II~\cite{Belle-II:2018jsg}, LHCb~\cite{Cerri:2018ypt} and future lepton colliders~\cite{Achasov:2023gey,Charm-TauFactory:2013cnj,CEPCStudyGroup:2018ghi,TLEPDesignStudyWorkingGroup:2013myl}, including the proposed CEPC and FCC-ee, which can also produce fruitful flavor results~\cite{Qin:2017aju,Ali:2018ifm,Cheng:2018khi,Ding:2019tqq,Qin:2020zlg,Shen:2022ffi}. 
Furthermore, the double-mixing CP asymmetry does not necessitate a non-zero strong phase, thereby circumventing strong-dynamics pollution in specific channels. Even when there is a non-zero strong phase, it turns out that the strong phase can potentially be determined directly from data without theoretical input, together with the weak phase. Therefore, by selecting appropriate channels, the double-mixing CP asymmetry can enable the extraction of CKM phase angles without hadronic uncertainties, making it sensitive to some dynamics beyond the SM.


In the rest of the paper, we will first present the general formulas for the double-mixing CP asymmetry in the process $M_1^0(t_1)\to \bar{M}_2^0(t_2)\to f$. We will then perform the numerical analysis of the $B^0_s \to \rho^0 \bar{K}^0 \to \rho^0 (\pi^-\ell^+{\nu}_\ell)$ decay channel, as an example to show that the double-mixing CP asymmetry can be very significant in numerics. Additionally, we will analyze the $B^0 \to D^0 {K} \to D^0 (\pi^+\ell^-\bar{\nu}_\ell)$ decay to exhibit that the involved weak phase can be extracted without any hadronic inputs. 





{\textit{Formulae.}} ---In the following derivation, we accept the convention that the mass eigenstates $M_{H,L}$ of the neutral mesons are superpositions of their flavor eigenstates 
\begin{eqnarray}
\ket{M_{H,L} } = p \ket{ M^0 } \mp q \ket{ \bar{M}^0 } \; ,
\end{eqnarray}
where $q,p$ are complex coefficients. The mass and decay width differences are defined as $\Delta m \equiv m_H - m_L$ and $\Delta \Gamma \equiv \Gamma_H-\Gamma_L$ such that the oscillation is formulated by 
\begin{eqnarray}
\ket{ M^0(t) } &=& g_+(t) \ket{M^0} - {q\over p} g_-(t) \ket{\bar{M}^0} \; ,  \\
\ket{ \bar{M}^0(t) } &=& g_+(t) \ket{\bar{M}^0} - {p\over q} g_-(t) \ket{{M}^0} \; , \nonumber \\ 
\text{with}\ g_{\pm}(t) & = & {1\over2} \left[ e^{-im_Ht-{1\over2}\Gamma_H t } \pm e^{ -im_L t-{1\over2}\Gamma_L t } \right] \; , \nonumber
\end{eqnarray}
in the case that the theory is CPT invariant. 

In this letter, we will focus on the cascade decay $M_1\to M_2\to f$ with the process happening via two oscillating paths $M_1^0\to \bar{M}_2^0 \to M_2^0$ and $M_1^0\to \bar{M}_1^0 \to M_2^0$, as shown in FIG.~\ref{fig:2chain}. A more comprehensive study including other cases is left for future~\cite{inpre}. The two-dimensional time-dependent CP asymmetry is defined by 
\begin{eqnarray}
A_\mathrm{CP}(t_1,t_2) &\equiv& {|\mathcal{M}|^2(t_1,t_2) - |\mathcal{\bar{M}}|^2 (t_1,t_2) \over |\mathcal{M}|^2(t_1,t_2) + |\mathcal{\bar{M}}|^2(t_1,t_2) }  \; ,
\end{eqnarray}
where the amplitude $\mathcal{M}(t_1,t_2)$ is the sum of amplitudes of the two paths $M_1^0\to \bar{M}_2^0 \to M_2^0$ and $M_1^0\to \bar{M}_1^0 \to M_2^0$, and the amplitude $\bar{\mathcal{M}}(t_1,t_2)$ is the CP conjugate of $\mathcal{M}(t_1,t_2)$. The decaying time $t_1$ and $t_2$ of $M_1$ and $M_2$ can be identified in experiments by using vertex detection techniques. To give prominence to the mixing effects, we assume that there are no direct CP asymmetries in the decay $M_1\to M_2$ or $M_2\to f$, {\it i.e.}, these decays are tree-amplitude dominant with vanishing or negligible penguin amplitudes.  Again, a more comprehensive analysis taking all these effects into consideration will be presented in an upcoming work~\cite{inpre}. 

{\bf Case 1.} We consider $M_1^0$ decaying into $\bar{M}_2^0$ associated with a CP eigenstate $f_{\rm CP}$ such as $\rho^0$. Then, the primary decay in the upper path $M_1^0\to \bar{M}_2^0 f_{CP}$ and the one in the lower path $\bar{M}_1^0\to {M}_2^0 f_{CP}$ are CP conjugates of each other. Thus, without direct CP violation we have $|\braket{\bar{M}_2^0}{M_1^0}| = |\braket{{M}_2^0}{\bar{M}_1^0}|$ and the decay amplitudes are related by $\braket{\bar{M}_2^0}{M_1^0} = \braket{{M}_2^0}{\bar{M}_1^0} e^{2i\omega}$, where $\omega$ is a pure weak phase. We further write the mixing parameters of $M_{1,2}$ as $(q/p)_{1,2} = |(q/p)_{1,2}| e^{-i\phi_{1,2}}$.  Then, the time-dependent CP asymmetry is calculated to be 
\begin{widetext}
\begin{eqnarray}\label{eq:acp}
A_\mathrm{CP}(t_1,t_2) &=&  {|g_{1,+}(t_1)|^2 C_+(t_2)  + |g_{1,-}(t_1)|^2 C_-(t_2) + e^{-\Gamma_1t_1}   \sinh{\Delta\Gamma_1t_1\over 2} \; S_h(t_2) + e^{-\Gamma_1t_1}   \sin{(\Delta m_1t_1)}\; S_n(t_2)
\over |g_{1,+}(t_1)|^2 C'_+(t_2) + |g_{1,-}(t_1)|^2 C'_-(t_2)   + e^{-\Gamma_1t_1}  \sinh{\Delta\Gamma_1t_1\over 2}\; S'_h(t_2)  + e^{-\Gamma_1t_1}  \sin{(\Delta m_1t_1)}\; S'_n(t_2) } \; ,
\end{eqnarray}
where 
\begin{eqnarray}\label{eq:c+c-}
C_+(t_2) &=& |g_{2,-}(t_2)|^2 \;  \left( \left| {p_2/ q_2} \right|^2 - \left| {q_2/ p_2} \right|^2 \right)   \;, \qquad C_-(t_2) = |g_{2,+}(t_2)|^2 \;  \left( \left| {q_1/ p_1} \right|^2 - \left| {p_1/ q_1} \right|^2 \right)  \; ,
\end{eqnarray}
consisting of $M_2$- and $M_1$-mixing induced CPV, respectively. The double-mixing CP asymmetry is reflected by the terms proportional to 
\begin{eqnarray}\label{eq:shn}
S_h(t_2) &=& {e^{-\Gamma_2t_2}\over2}  [ - 2 \sin{(\Delta m_2t_2)}  \sin(\phi_1+\phi_2+2\omega )   
+ \sinh{\Delta\Gamma_2t_2\over2} \left( \left|{q_1 \over p_1}\right|\left|{p_2\over q_2}\right| - \left|{p_1\over q_1}\right| \left|{q_2\over p_2}\right| \right) 
\cos(\phi_1+\phi_2+2\omega )  ] \;,  \nonumber \\
S_n(t_2) &=& {e^{-\Gamma_2t_2}\over2}  [   2 \sinh{\Delta\Gamma_2t_2\over2}  \sin(\phi_1+\phi_2+2\omega ) 
+\sin{(\Delta m_2t_2)} \left( \left|{q_1 \over p_1}\right|\left|{p_2\over q_2}\right| - \left|{p_1\over q_1}\right| \left|{q_2\over p_2}\right| \right)  
\cos(\phi_1+\phi_2+2\omega ) ] \;,  
\end{eqnarray}
\end{widetext}
where some doubly suppressed small quantities are neglected. The phase angle dependence on $\phi_1+\phi_2$ clearly indicates that they are induced by the inference between the $M_1$-oscillating path and the $M_2$-oscillating path. The $S_h$ and $S_n$ terms have very different time dependence: $S_n$ has sine dependence on $t_1$ and hyperbolic sine dependence on $t_2$, and $S_h$ has hyperbolic sine dependence on $t_1$ and sine dependence on $t_2$. The two different types of time dependence can be used to separate $S_h$ and $S_n$ in $A_\mathrm{CP}(t_1,t_2)$ and analyze the physical implications of each. It can also be observed that in the absence of strong phases in decays and mixings, {\it i.e.}, $|\braket{\bar{M}_2^0}{M_1^0}| = |\braket{{M}_2^0}{\bar{M}_1^0}|$, and $|q/p|=1$, the double-mixing CP asymmetries $S_h$ and $S_n$ still exist, and it thus provides a clean environment to determine CKM phases.


The terms contributing to the denominator of \eqref{eq:acp} are given by 
\begin{eqnarray}\label{eq:deno}
C'_+(t_2) &=& 2|g_{2,-}(t_2)|^2   \;, \nonumber \\
C'_-(t_2) &=& 2 |g_{2,+}(t_2)|^2   \; , \nonumber \\
S'_h(t_2) & = & {e^{-\Gamma_2t_2}}   \sinh{\Delta\Gamma_2t_2\over2} \cos(\phi_1+\phi_2+2\omega )  \; , \nonumber \\
S'_n(t_2) & = & {e^{-\Gamma_2t_2}}   \sin{(\Delta m_2t_2)} \cos(\phi_1+\phi_2+2\omega )  \; ,
\end{eqnarray}
with suppressed terms neglected. 

{\bf Case 2.} Another case is that the particles $p$ produced in the primary decay $M_1\to M_2 p$ is not a CP eigenstate, {\it e.g.,} $D^0$. Then, the involved decay $M_1^0\to \bar{M}_2^0 p$ in the upper path shown by FIG.~\ref{fig:2chain}, and the corresponding decay in the lower path $\bar{M}_1^0\to {M}_2^0 p$ are not the CP conjugations of each other. Therefore, more parameters are necessary to formulate the relations between the primary decay amplitudes in the two paths and their charge conjugations, 
\begin{eqnarray}\label{eq:case2}
{A(\bar{M}_1^0\to p M_2^0)/ A(M_1^0\to \bar{p} \bar{M}_2^0)} &=& e^{-2i\omega_1} \; ,\nonumber \\
{A(M_1^0\to p \bar{M}_2^0)/A(M_1^0\to \bar{p} \bar{M}_2^0)} &=& r e^{i(\delta +\omega_2)} \; ,  
\end{eqnarray}
where $\omega_{1,2}$ are the weak phases, $r$ is the magnitude ratio, and $\delta$ is the strong phase. 

The two-dimensional time-dependent CP asymmetry is calculated to be 
\begin{widetext}
\ 
\begin{eqnarray}\label{eq:case2}
A_\mathrm{CP}(t_1,t_2) &=&  {  e^{-\Gamma_1t_1}   \sinh{\Delta\Gamma_1t_1\over 2} \; S_h(t_2) + e^{-\Gamma_1t_1}   \sin{(\Delta m_1t_1)}\; S_n(t_2)
\over |g_{1,+}(t_1)|^2 r^2 C'_+(t_2) + |g_{1,-}(t_1)|^2 C'_-(t_2)   + e^{-\Gamma_1t_1}  \sinh{\Delta\Gamma_1t_1\over 2}\; S'_h(t_2)  + e^{-\Gamma_1t_1}  \sin{(\Delta m_1t_1)}\; S'_n(t_2) } \; ,
\\
S_h(t_2) &=&  - e^{-\Gamma_2t_2}\; r \sin \omega'  [  \sin\delta \sinh{\Delta\Gamma_K\over2}t_2 + \cos\delta \sin{\Delta m_Kt_2}  ] \; ,\nonumber \\
S_n(t_2) &=& e^{-\Gamma_2t_2}\; r \sin \omega'  [   \cos\delta \sinh{\Delta\Gamma_K\over2}t_2 -  \sin\delta \sin{\Delta m_Kt_2}  ] \; ,\nonumber \\
S'_h(t_2) &=&   e^{-\Gamma_2t_2}\;r  \cos \omega' [  \cos\delta \sinh{\Delta\Gamma_K\over2}t_2 -  \sin\delta \sin{\Delta m_Kt_2}  ] \;, \nonumber \\
S'_n(t_2) &=&  e^{-\Gamma_2t_2}\; r  \cos \omega'  [   \sin\delta \sinh{\Delta\Gamma_K\over2}t_2 +  \cos\delta \sin{\Delta m_Kt_2}  ] \;. \nonumber 
\end{eqnarray}

\end{widetext}
where the weak phase $\omega' = 2\omega_1 + \omega_2 + \phi_1 + \phi_2$, and the functions $C'_{\pm}$ take the same form as in \eqref{eq:deno}. In the calculation, we have neglected the CP violation in mixing, {\it i.e.}, setting $|q_1/p_1|= |q_2/p_2|=1$.

For the decay chain constructed by the oscillation paths $M_1^0\to M_2^0 \to \bar{M}_2^0$ and $M_1^0\to \bar{M}_1^0 \to \bar{M}_2^0$, the results can be obtained by taking \eqref{eq:acp} and \eqref{eq:case2} and replacing $q_2/p_2$ with $p_2/q_2$, {\it i.e.}, $|q_2/p_2| \to |p_2/q_2|$ and $\phi_2\to -\phi_2$.

{\it Phenomenology} --- We perform the phenomenological analysis of two decay channels, as the examples corresponding to the two cases above. For case 1, $B^0_s \to \rho^0 K \to \rho^0 (\pi^-\ell^+{\nu}_\ell)$ is analyzed, showing that the value of the double-mixing CP asymmetry in this channel is very significant, in the absence of strong phases. For case 2, $B^0 \to D^0 {K} \to D^0 (\pi^+\ell^-\bar{\nu}_\ell)$ is analyzed, showing that the strong phase and the weak phase can be simultaneously determined by measurements of the double-mixing CP asymmetry. 

{\bf Example channel 1.}
The $B^0_s \to \rho^0 K \to \rho^0 (\pi^-\ell^+{\nu}_\ell)$ decay channel is taken as the first example. The process has $B^0_s \to \rho^0 \bar{K}^0 \to \rho^0 K^0 \to \rho^0 \pi^-\ell^+{\nu}_\ell$ and $B^0_s \to \bar{B}^0_s \to \rho^0 K^0 \to \rho^0 \pi^-\ell^+{\nu}_\ell$ as the two paths shown in FIG.~\ref{fig:2chain}. Because both $B^0_s$ and $K^0$ have large mixing effects, their interference could also be large. 

As the penguin operator contributions to $B^0_s \to \rho^0 \bar{K}^0$ are suppressed due to small Wilson coefficients~\cite{Beneke:2003zv,Ali:2007ff}, we neglect them for simplicity, {\it i.e.}, the direct CP asymmetry is neglected. Then, the decay amplitude $\braket{\rho^0 \bar{K}^0}{B^0_s}$ and its charge conjugation $\braket{\rho^0 K^0}{\bar{B}^0_s}$ share the same magnitude, and the phase difference is given by $e^{i2\omega} = -{(V_{ub}^*V_{ud})/ ( V_{ub}V_{ud}^*)}$, where the minus sign is caused by $CP|\rho^0 K^0\rangle = - |\rho^0 \bar{K}^0\rangle$ in a pseudoscalar meson decay. We also neglect the indirect CP asymmetries induced by $B_s$ and $K$ mixings, {\it i.e.,} taking $|q/p|=1$, and then the phases of the mixing coefficients are approximately given by $e^{-i\phi_1} = e^{-i\phi_{B_s}} = {(V_{tb}^*V_{ts})/ (V_{tb}V_{ts}^*})$ and $e^{-i\phi_2} =e^{-i\phi_K} = {(V_{cd}^*V_{cs})/(V_{cb}V_{cs}^*)}$. With these approximations, the only non-vanishing CP asymmetry is induced by the double-mixing interference, contained in $S_h$ and $S_n$ given by \eqref{eq:shn}.

With the numerical inputs listed in TABLE~\ref{table:parameters}, where the neutral meson average decay width is defined by $\Gamma_M\equiv (\Gamma_{M,H}+\Gamma_{M,L})/2$, we calculate the numerical results for the two-dimensional time-dependent CP violation observable $A_{\rm CP}(t_1,t_2)$ \eqref{eq:acp}, and the contributions $A_{h}(t_1,t_2)$ and $A_{n}(t_1,t_2)$ proportional to the $S_h$ and $S_n$ terms \eqref{eq:shn}, respectively. The result for the $A_{\rm CP}(t_1,t_2)$ dependence on $t_1$ and $t_2$ is displayed in the left panel of FIG.~\ref{fig:BsrhoK}. It can be observed that the magnitude of the peak values can exceed 50\%. If the data sample is not large enough for a two-dimensional time-dependence analysis, we can also integrate out one time dimension and get the evolution along the remaining one. Integrating $t_2$ from 0 to the Kaon average lifetime $\tau_{K}\equiv 1/\Gamma_{K}$, we obtain the $t_1$ dependence of $A_{\rm CP}$ displayed in the middle panel of FIG.~\ref{fig:BsrhoK}; integrating $t_1$ from $\tau_{B_s}$ to $5\tau_{B_s}$ with $\tau_{B_s}\equiv 1/\Gamma_{B_s}$, we obtain the $t_2$ dependence of $A_{\rm CP}$ displayed in the right panel of FIG.~\ref{fig:BsrhoK}. The contributions $A_{h}$ and $A_{n}$ are displayed by dashed and dotted curves, respectively. The time evolution along $t_2$ is dominated by $A_h$, because $A_n$ highly oscillates along the integrated time $t_1$. The time evolution along $t_1$ is dominated by $A_n$, but the $A_h$ contribution is also considerable.

\begin{widetext}

\begin{table}[!h]
\caption{The input parameters and their values, with $x_M\equiv \Delta m_M/\Gamma_M$ and $y_M\equiv \Delta \Gamma_M/(2\Gamma_M)$, respectively.}\label{table:parameters}
{
\begin{tabular}{cr|cr}
\hline
 \textbf{Parameter}    &  \textbf{Value} &  \textbf{Parameter}    &  \textbf{Value} \\
\hline
  $\phi_{B_s}$      &    $(-2.106\pm 0.135)^\circ$~\cite{ParticleDataGroup:2022pth} &  $\phi_{B}$      &    $(44.4 \pm 1.4)^\circ$~\cite{HFLAV:2022pwe}  \\
  $x_{B_s}$   &   $27.01\pm 0.10$~\cite{ParticleDataGroup:2022pth} &
  $x_{B}$   &   $0.769\pm 0.004$~\cite{ParticleDataGroup:2022pth} \\
  $y_{B_s}$     &   $-0.064\pm 0.003$~\cite{ParticleDataGroup:2022pth} &
  $y_{B}$     &   $-0.0005\pm 0.0050$~\cite{ParticleDataGroup:2022pth} \\
  $\phi_K$      &    $(0.176\pm 0.001)^\circ$~\cite{ParticleDataGroup:2022pth} &
  $2\omega$     &    $(-48.907\pm 3.094)^\circ$~\cite{ParticleDataGroup:2022pth} \\
  $x_K$         &    $0.946\pm 0.002$~\cite{ParticleDataGroup:2022pth} &
  $2\omega_1$     &    $0^\circ$~\cite{ParticleDataGroup:2022pth} \\
  $y_K$    &    $-0.996506\pm 0.000016$~\cite{ParticleDataGroup:2022pth} &
  $\omega_2$     &    $\pqty{65.54\pm 1.55}^\circ$~\cite{ParticleDataGroup:2022pth} \\
\hline
\end{tabular}
}
\end{table}

\begin{figure}[htbp]
    \centering
    \includegraphics[keepaspectratio,width=6.cm]{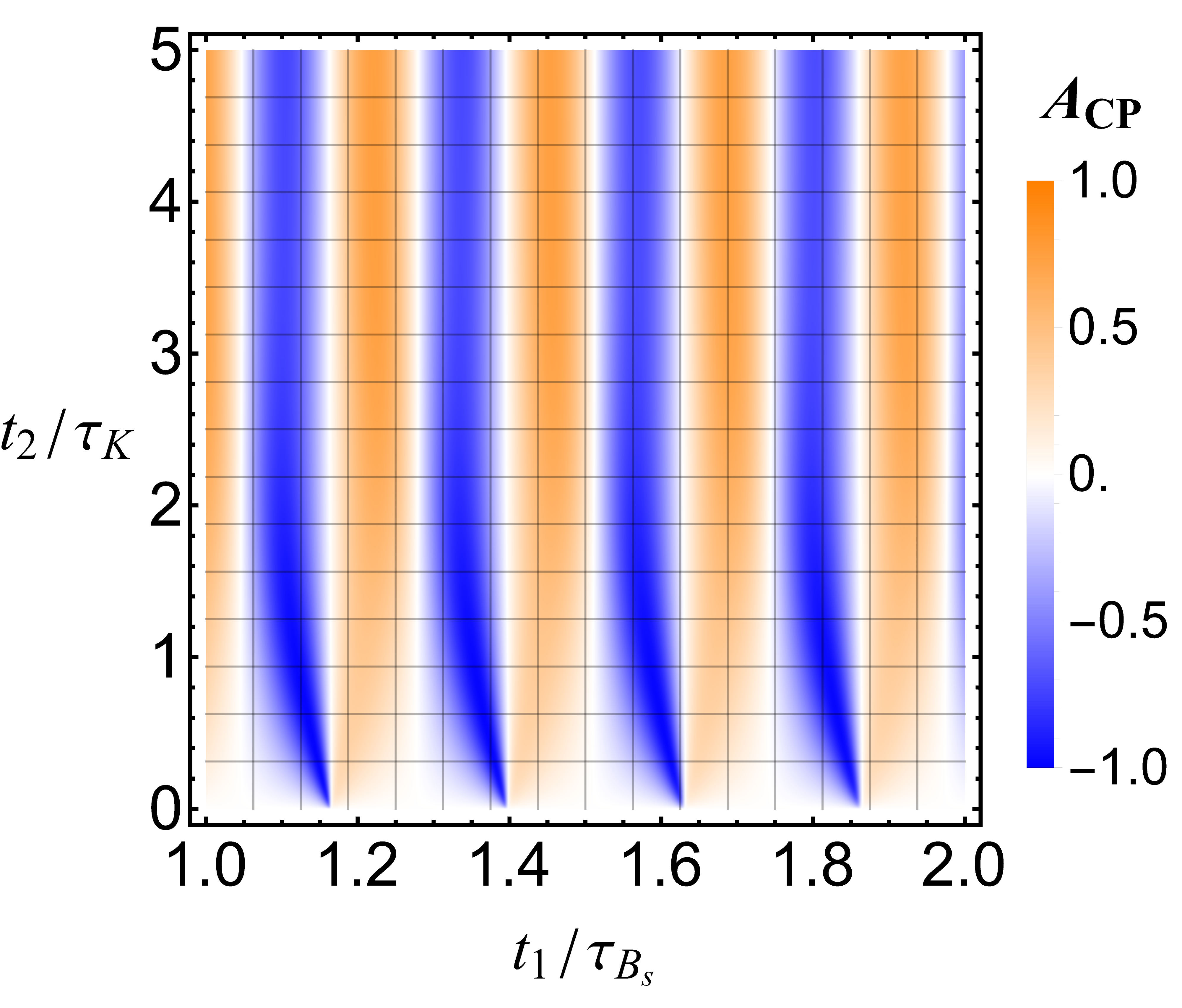}
    \hspace{0.8cm}
    \includegraphics[keepaspectratio,width=4.8cm]{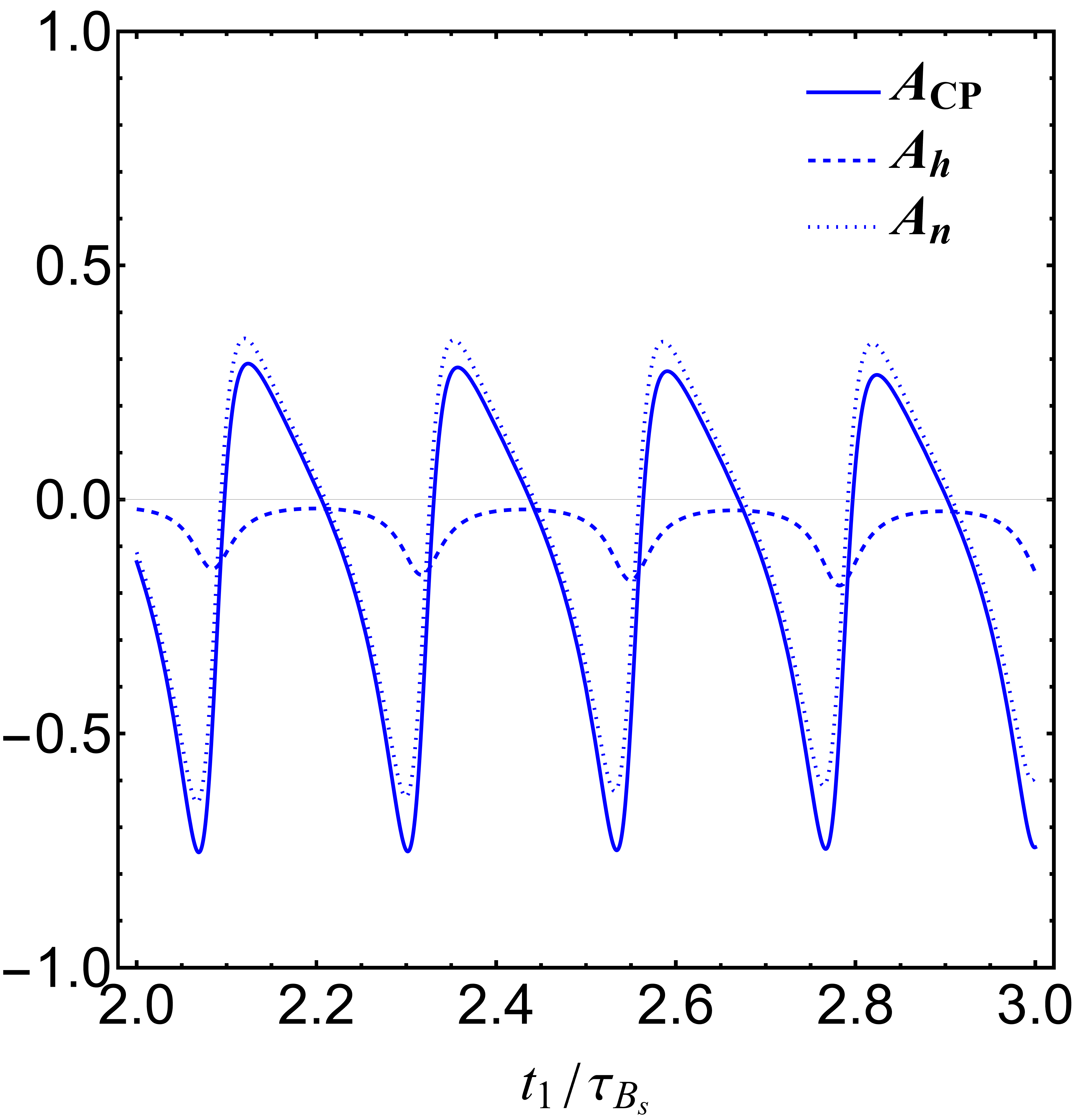}
     \hspace{0.8cm}
    \includegraphics[keepaspectratio,width=4.7cm]{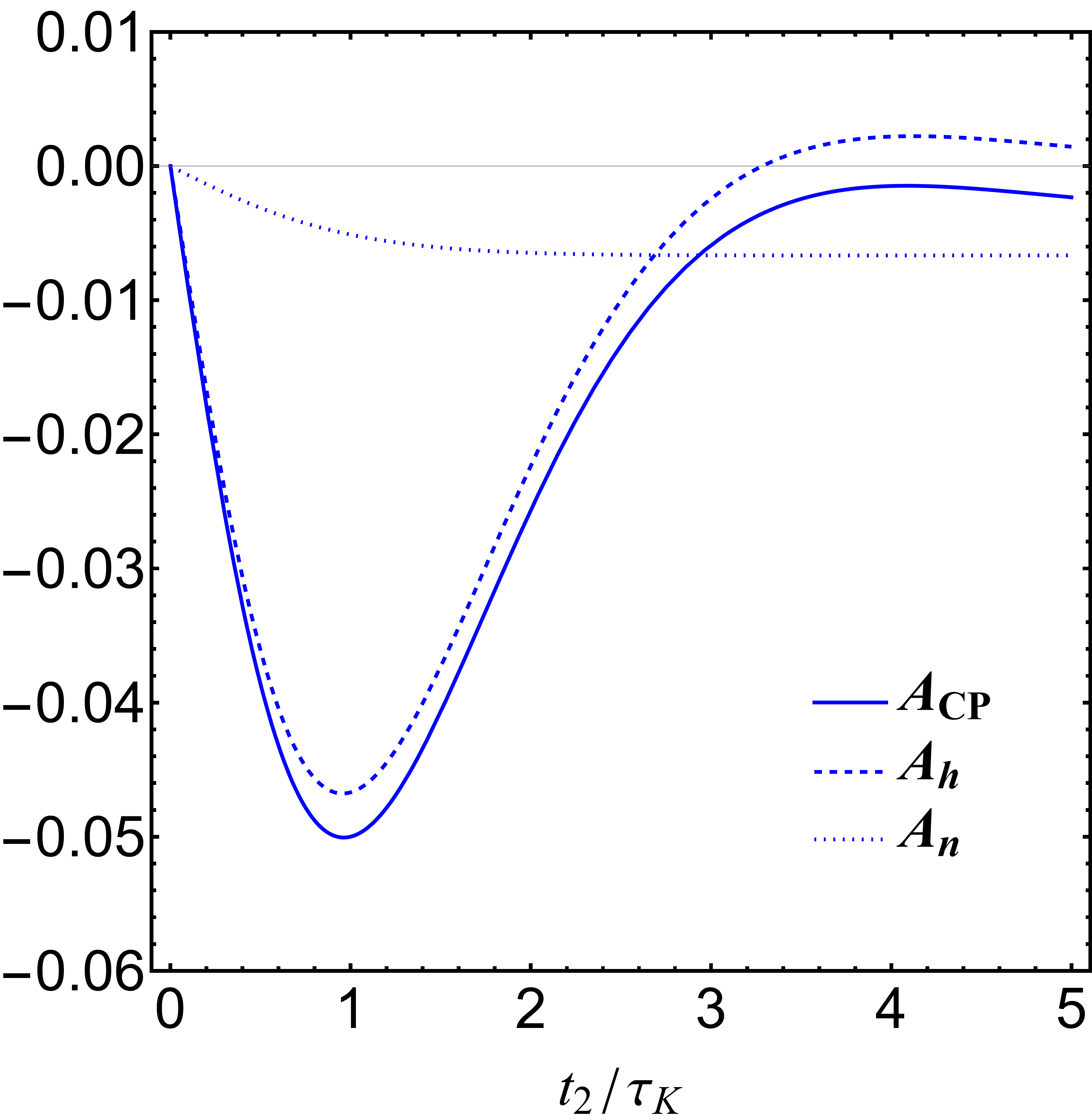}
    \caption{Time dependence of the double-mixing CP asymmetry $A_{\rm CP}$ in $B^0_s(t_1) \to \rho^0 K(t_2) \to \rho^0 (\pi^-\ell^+{\nu}_\ell)$. The left panel displays the two-dimensional time dependence. The middle panel and the right panel display the dependence on $t_1$ (with $t_2$ integrated from 0 to $\tau_{K}$) and $t_2$ (with $t_1$ integrated from $\tau_{B_s}$ to $5\tau_{B_s}$), respectively. }\label{fig:BsrhoK}
\end{figure}
\end{widetext}

{\bf Example channel 2.} ---
The $B^0 \to D^0 {K} \to D^0 (\pi^+\ell^-\bar{\nu}_\ell)$ decay channel is taken as the second example, showing how CKM phases can be determined by measurements of the double-mixing CP asymmetry. Although a strong phase is involved in this channel, it can be extracted together with the weak phase from data without any theoretical input needed. 

\begin{widetext}

\begin{figure}[htbp]
    \centering
    \includegraphics[keepaspectratio,width=5.5cm]{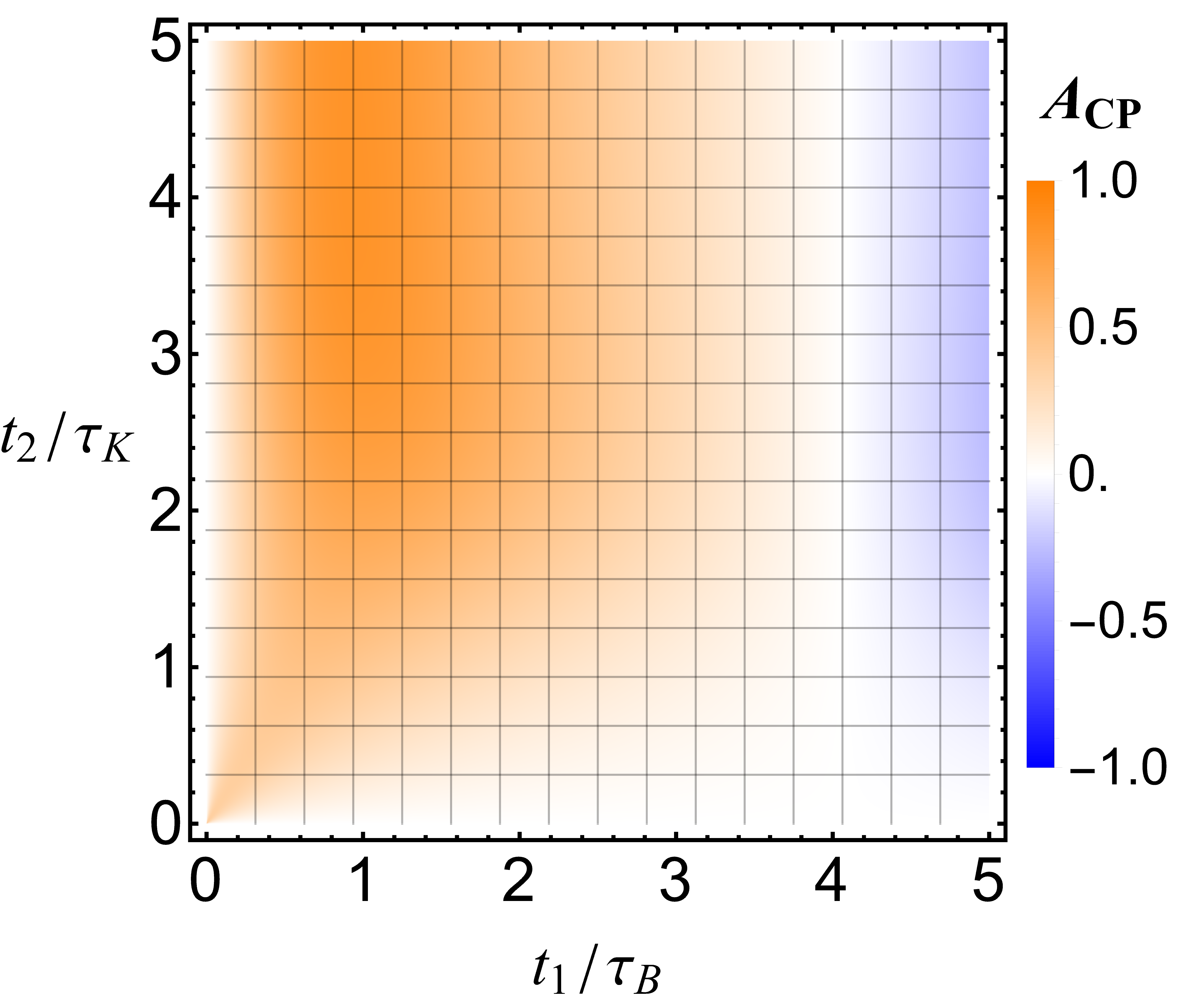}
    \hspace{0.8cm}
    \includegraphics[keepaspectratio,width=4.4cm]{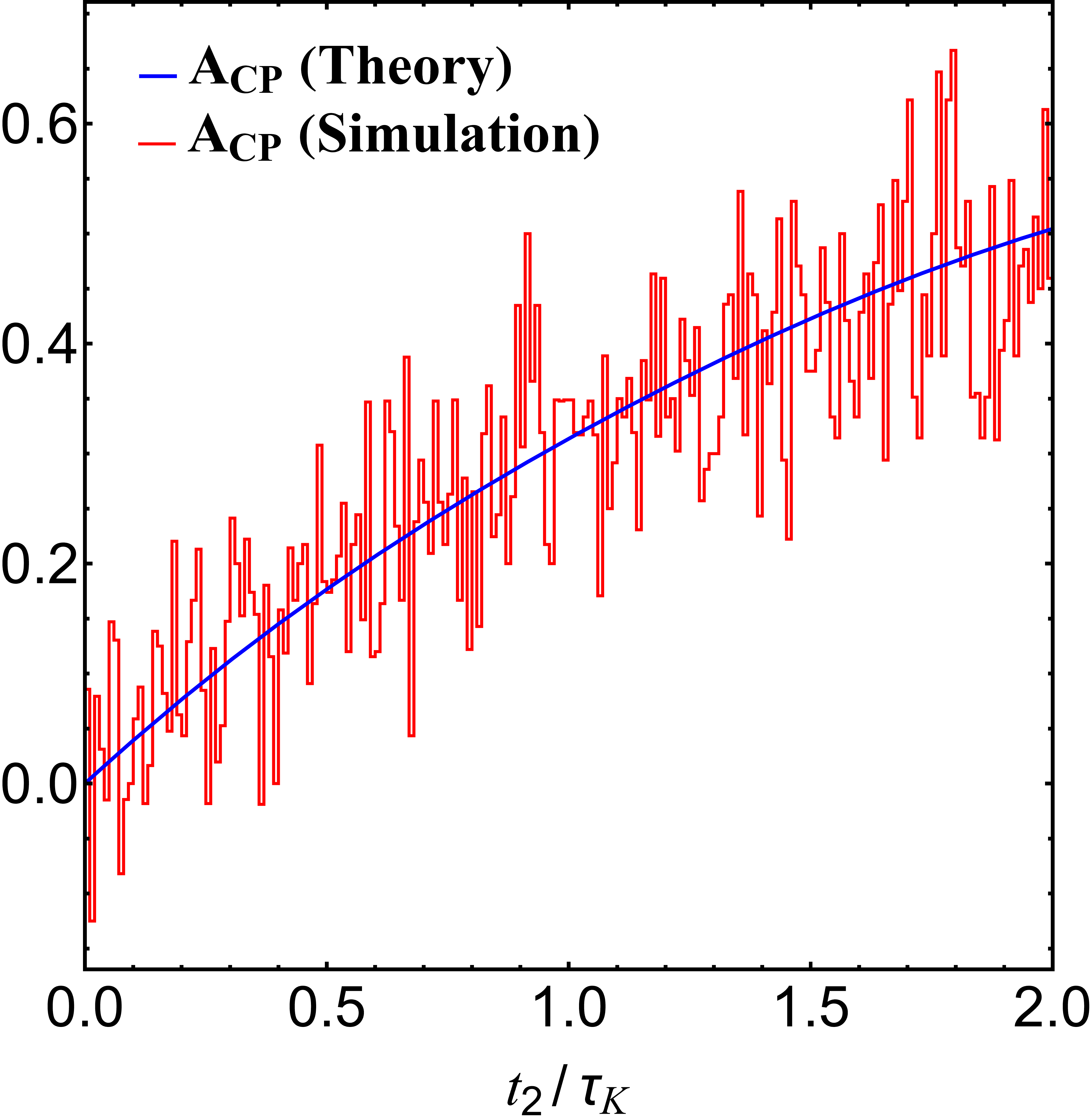}
     \hspace{0.8cm}
    \includegraphics[keepaspectratio,width=4.9cm]{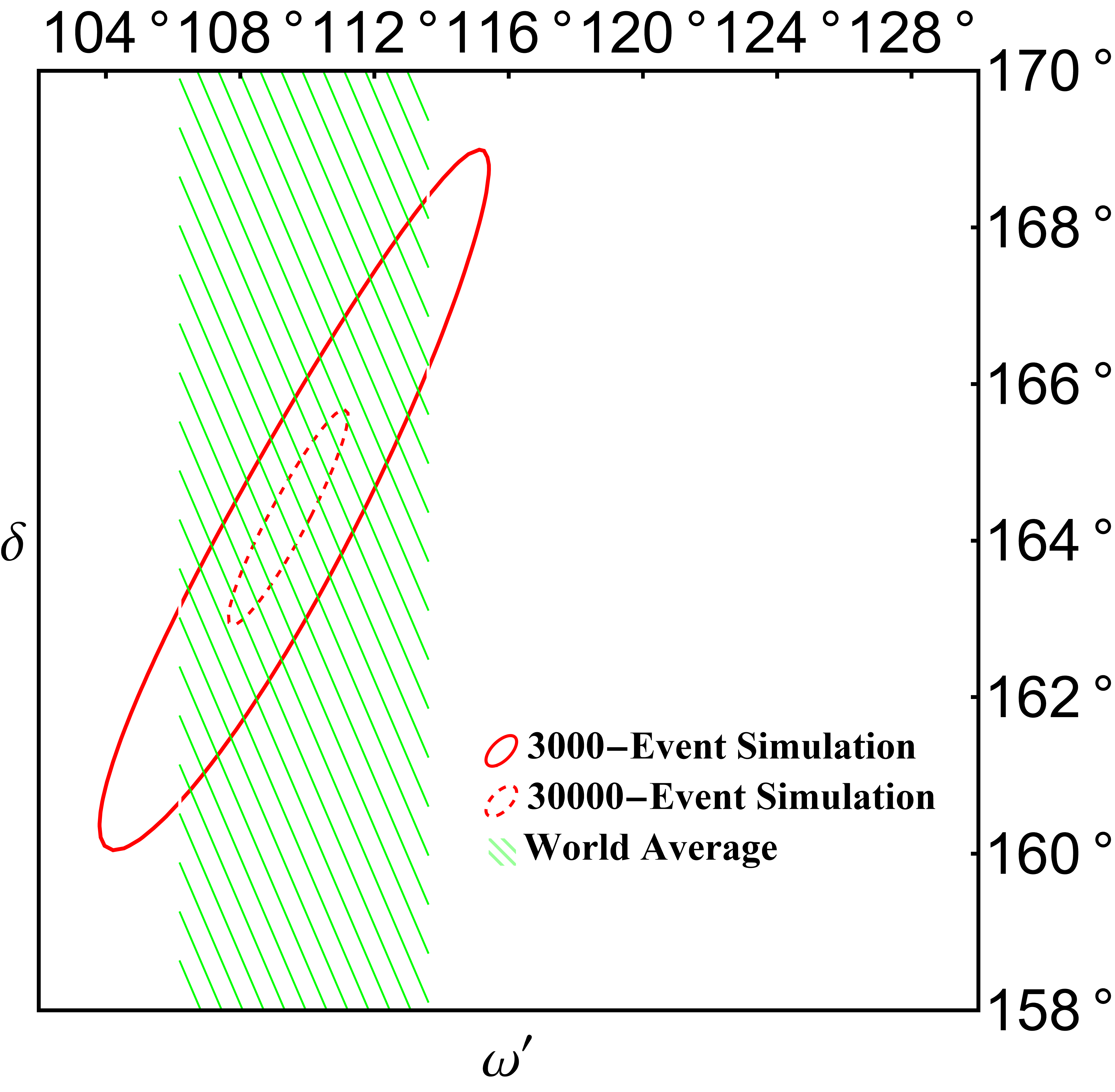}
    \caption{Numerical results for the double-mixing CP asymmetry $A_{\rm CP}$ in $B^0(t_1) \to D^0 {K}(t_2) \to D^0 (\pi^+\ell^-\bar{\nu}_\ell)$ and the extraction of the weak and strong phases from its numerical simulation. The left panel displays the two-dimensional time dependence of $A_{\rm CP}$. The middle panel displays the $t_2$ dependence of $A_{\rm CP}$  (with $t_1$ integrated from 0 to $3\tau_{B}$) and the corresponding simulated result. The right panel presents the strong phase $\delta$ and weak phase $\omega'$ at 68\% confidence level determined by the simulated data samples, with 3000 and 30000 simulated events, respectively, compared to the world average for the weak phase $\omega'$ at 68\% confidence level.}\label{fig:BDK}
\end{figure}
\end{widetext}

The $B^0 \to D^0 {K} \to D^0 (\pi^+\ell^-\bar{\nu}_\ell)$ channel has two oscillating paths, $B^0 \to D^0 K^0 \to D^0\bar{K}^0$ and $B^0 \to \bar{B}^0 \to D^0\bar{K}^0$, which have a nonzero relative strong phase between each other, as parametrized in \eqref{eq:case2}. Comparing to the setup in {\bf Case 2}, we have $p= D^0$ and $\bar{M}_2^0 = K^0$. To apply \eqref{eq:case2}, we need to additionally flip the sign of $\phi_2$, {\it i.e.}, the total weak phase is $\omega' = 2\omega_1 + \omega_2 + \phi_1 -\phi_2$, with 
\begin{eqnarray}
&&2\omega_1 = \arg {V_{cb}^*V_{us}\over V_{cb} V_{us}^*} \; , \; 
\omega_2 = \arg {V_{ub}^* V_{cs}\over V_{cb}^*V_{us}} \; , \nonumber \\
&& \phi_1  = \phi_B = - \arg{V_{tb}^*V_{td}\over V_{tb}V_{td}^*}\; ,  \; \phi_2 = \phi_K \; . 
\end{eqnarray}
It can be checked that $\omega'$ is approximately $2\beta+\gamma$ in the conventional parametrization of the CKM phase angles~\cite{ParticleDataGroup:2022pth}.
With the values of the input parameters listed in TABLE~\ref{table:parameters}, and choosing the reference values for the ratio $r=0.366$ and the strong phase $\delta=164^\circ$ (close to the measured values of the corresponding parameters for $B^0 \to D K^{*}$~\cite{LHCb:2016aix,LHCb:2016bxi}), we obtain the numerical result for the two-dimensional time-dependent CP asymmetry $A_{\rm CP}(t_1,t_2)$ and display it in the left panel of FIG.~\ref{fig:BDK}. Integrating out $t_1$ from 0 to 3$\tau_B$, the $t_2$ dependence of $A_{\rm CP}$ is retained, as shown in the middle panel of FIG.~\ref{fig:BDK}. It is observed that in both cases the CP violation effects are considerable.

To show how the weak phase is extracted from measurements of the double-mixing CP asymmetry of this channel, we simulate the events for both $B^0 \to D^0 {K} \to D^0 (\pi^+\ell^-\bar{\nu}_\ell)$ and its charge-conjugate process. Assuming that the semileptonic decays of Kaons within 2$\tau_{K_S}$ in the rest frame can be perfectly detected by Belle II, we simulate 3000 $B^0 \to D^0 {K} \to D^0 (\pi^+\ell^-\bar{\nu}_\ell)$ events and the corresponding charge conjugate events. The simulated CP asymmetry is shown by the red histogram in the middle panel of FIG.~\ref{fig:BDK}. Afterwards, the ratio $r$ and the strong and weak phases $\delta$ and $\omega'$ are treated as unknown parameters to be determined. By fitting the formulas for the double-mixing CP asymmetry \eqref{eq:case2} and the CP-averaged branching ratio to the simulated events, the three parameters are extracted to be 
\begin{eqnarray}
r &=& 0.367\pm 0.014 \;, \; \delta =(164.1\pm 4.1)^\circ \;, \nonumber \\ 
\omega' &=& (108.9\pm 4.8)^\circ \; , 
\end{eqnarray}
where only the statistical uncertainties are considered. The 68\% confidence interval for $\delta-\omega'$ is shown in the right panel of FIG.~\ref{fig:BDK} by the red solid contour. The precision of the weak phase $\omega'\approx 2\beta +\gamma$ is comparable to the world average value of all the current experiments, $(109.9 \pm3.7)^\circ$. If 10 times events are collected, the precision can be further improved by about 3 times, as shown by the red dashed contour in the right panel of FIG.~\ref{fig:BDK}. 

{\it Conclusion.} --- In conclusion, we have discovered a new type of CP violation effect, the double-mixing CP asymmetry, which exists in cascade decays involving two neutral mesons oscillating. This effect is dependent on two time variables, allowing for a two-dimensional time dependence analysis. Unlike direct CP asymmetries, the existence of the double-mixing CP asymmetry does not require a non-zero strong phase. Even in the presence of a strong phase, it can be directly extracted from data with the corresponding weak phase in appropriate channels. As a result, the double-mixing CP asymmetry provides a means to directly extract weak phases without any pollution from strong dynamics, which is crucial for CKM matrix determination and new physics search. Furthermore, the double-mixing CP asymmetries can be numerically significant in certain channels, making them very promising for experimental measurement.

{\it Acknowledgement.} --- The authors are grateful to Cai-Dian L\"u, Wen-Bin Qian, Liang Sun and Yue-Hong Xie for useful discussions.
This work is supported by Natural Science Foundation of China under grant No. 12005068.

\end{document}